\documentclass[twocolumn,showpacs,preprintnumbers,
floatfix,amsmath,amssymb,eqsecnum]{revtex4}% Physical Review B

\usepackage{epsfig,psfrag}
\usepackage{dcolumn}     % Align table columns on decimal point
\usepackage{bm}          % bold math

\begin{document}

%\preprint{preprint not for distribution}

\title{Effective low-energy theory of
superconductivity in carbon nanotube ropes}
\author{A.~De~Martino and R.~Egger}

\affiliation{Institut f\"ur Theoretische Physik, 
Heinrich-Heine-Universit\"at,
 D-40225 D\"usseldorf, Germany}

\date{\today}

\begin{abstract}
We derive and analyze the low-energy theory of superconductivity in
carbon nanotube ropes.  A rope is modelled as an
array of metallic nanotubes, taking into account 
phonon-mediated as well as Coulomb interactions, and 
arbitrary Cooper pair hopping amplitudes
(Josephson couplings) between different tubes.
We use a systematic cumulant expansion to construct the
Ginzburg-Landau action including quantum fluctuations. 
The regime of validity is carefully established, and the 
effect of phase slips is assessed.
Quantum phase slips are shown to cause a depression of the critical 
temperature $T_c$ below the mean-field value, and a 
temperature-dependent resistance below $T_c$.  
We compare our theoretical results to recent experimental data
of Kasumov {\sl et al.} [Phys. Rev. B {\bf 68}, 214521 (2003)] for
the sub-$T_c$ resistance,
and find good agreement with only one free fit parameter.
Ropes of nanotubes therefore represent superconductors in the one-dimensional
few-channel limit.
\end{abstract}

\pacs{73.63.Fg, 74.78.Na, 74.25.Fy}

\maketitle

\section{Introduction}

Over the past decade, the unique mechanical, electrical,
and optical properties of carbon nanotubes, 
including the potential for useful technological applications,
have created a lot of excitement \cite{nts,nts2}.
While many of these properties are well understood
by now, the experimental observation of intrinsic 
\cite{kociak,kasnew,tang} and anomalously strong proximity-induced 
\cite{morpurgo,kasumov} superconductivity continues to pose open
questions to theoretical understanding. 
In this paper we present a theory of  
one-dimensional (1D) superconductivity as found in ropes of
carbon nanotubes \cite{kociak,kasnew} and potentially in other
nanowires.  Ropes are 1D materials in the sense that there is
only a relatively small number of
propagating channels (typically, $N\approx 10$ to $100$) 
available to electronic transport.
While most other 1D materials tend to become insulating at low
temperatures due to the Peierls transition or as a consequence
of electron-electron interactions, nanotubes can stay metallic
down to very low temperatures \cite{nts}.  
If the repulsive electron-electron interactions can be 
overcome by attractive phonon-mediated interactions, 
ropes of nanotubes can then exhibit a superconducting transition.

However, due to strong 1D fluctuations, this transition is 
presumably rather broad, and the question of how precisely
superconductivity breaks down as the number of propagating channels
decreases has to be answered by theory.   Experimentally, the
breakdown of superconductivity manifests itself 
as a temperature-dependent
resistance below the transition temperature $T_c$, which 
becomes more and more pronounced as the rope gets thinner \cite{kasnew}.
According to our theory, this resistance is caused
by quantum phase slips, and therefore the experimental data 
published in Ref.~\cite{kasnew} have in fact explored a 
regime of 1D superconductivity with clear evidence for quantum
phase slip events that had not been reached before.
To the best of our knowledge, nanotube ropes represent wires
with the smallest number of propagating channels 
showing intrinsic superconductivity, even when
compared to the amorphous MoGe wires of diameter
$\approx 10$~nm studied in Ref.~\cite{lau}, 
where still several thousand channels are available.

We theoretically analyze superconductivity in nanotube ropes 
by starting from the microscopic model of an array of
$N$ individual metallic single-wall nanotubes (SWNTs) without disorder, 
with effectively attractive on-tube interactions 
and inter-tube Josephson couplings.  A similar model has been
suggested by Gonz{\'a}lez \cite{gonzalez1,gonzalez2}.
In the absence of the Josephson couplings, each SWNT would then
correspond to a {\sl Luttinger liquid}\
 with interaction parameter $g_{c+}>1$,
where $g_{c+}=1$ marks the noninteracting limit. For simplicity, we
take the same $g_{c+}$ on each SWNT.
For example, for $(10,10)$ armchair SWNTs, assuming good screening
of the repulsive Coulomb interactions, phonon exchange via a breathing
mode (as well as optical phonon modes) leads to $g_{c+}\approx 1.3$,
see Ref.~\cite{ademarti}.  In the case of attractive interactions,
the dominant coupling mechanism between different SWNTs is then given
by Cooper pair hopping, while
single-particle hopping is drastically  suppressed by 
momentum conservation arguments \cite{kane,gonzalez1}.
The coupling among different SWNTs is thus encoded in a
{\sl Josephson coupling matrix} $\Lambda_{ij}$, where $i,j=1,\ldots,N$.
As different nanotube chiralities are randomly distributed in 
a rope, only $1/3$ of the SWNTs can be expected to be metallic. 
In general, the $\Lambda_{ij}$ matrix should therefore be
drawn from an appropriate random distribution.  We consider 
below one individual rope with a fixed (but unspecified) 
matrix, and derive general statements valid for
arbitrary $\Lambda_{ij}$.  In that sense, our theory allows to
capture some disorder effects, at least qualitatively.   However,
since typical elastic mean free paths in SWNTs exceed $1 \mu$m 
\cite{nts}, disorder effects within individual SWNTs are ignored completely. 
The above reasoning leads us to the problem of 
$N$ coupled strongly correlated Luttinger liquids, where 
the number of ``active'' chains $N \lesssim 100$ with 
reference to the experiments of Ref.~\cite{kasnew}.
This is a difficult problem that neither permits 
the use of classical Ginzburg-Landau (GL) theory
nor of the standard BCS approach,
in contrast to the situation encountered in,
e.g., wide quasi-1D organic superconductors \cite{schulz}.

The approach taken in this paper is sketched next.
After a careful derivation of the coupled-chain action in Sec.~\ref{sec2}, we
proceed by introducing the appropriate order parameter
field.  In Sec.~\ref{sec3}, we then perform a cumulant expansion in this order
parameter, and thereby give a 
microscopic derivation of the quantum GL action,  which 
then allows to make further progress.  We establish the
temperature regime where this theory is reliable, and
then focus on the important phase fluctuations of the
order parameter field. At temperatures $T$ well below a
mean-field transition temperature $T_c^0$, amplitude fluctuations
are shown to be massive, and hence the amplitude can safely 
be treated in mean-field theory.  The massless phase fluctuations
then capture the important physics, and  
we specify the resulting effective low-energy action, valid at 
temperatures well below $T_c^0$.
Based on this action, Sec.~\ref{sec4} explains why
quantum phase slips (QPSs) 
\cite{tinkham,zaikin1,zaikin2,blatter} are 
crucial for an understanding of the experimental results of 
Refs.~\cite{kociak,kasnew}.  First, they cause a depression of the
transition temperature $T_c$ below the mean-field critical 
temperature  $T_c^0$.  
Furthermore, for $T<T_c$, a finite resistance $R(T)$ due to QPSs appears, 
which exhibits approximate power-law scaling.
We determine the full temperature dependence 
of $R(T<T_c)$ for arbitrary rope length in Sec.~\ref{sec5}.
In Sec.~\ref{sec6}, we then compare these results for $R(T)$
to the experimental data of Ref.~\cite{kasnew}, focussing on two
of their samples.  Finally, Sec.~\ref{sec7} offers some concluding
remarks.   Throughout the paper, we put $\hbar=k_B =1$.

\section{Model and order parameter} \label{sec2}

We consider a rope consisting of $N$
metallic SWNTs participating in superconductivity.  
Experimentally, this number can be found from 
the residual resistance measured as offset in the
resistance when extrapolating down to $T=0$ 
\cite{kasnew}.  Due to the attached  normal electrodes
in any two-terminal measurement of the rope,
despite of the presence of superconductivity, there will always be 
a finite contact resistance $R_c$. 
Since each metallic tube contributes two conduction channels, 
assuming good transparency for the contacts between metallic tubes
and the electrodes,
this is given by 
\begin{equation}\label{contactres}
R_c= \frac{R_Q}{2N}, \quad R_Q=h/2e^2\simeq 12.9\, k\Omega.
\end{equation}
Extrapolation of experimental data for the resistance $R(T)$ down
to $T\to 0$ within the
superconducting regime then allows to measure $R_c$, and hence $N$.
Good transparency of the contacts is warranted by the sputtering technique
used to fabricate and contact the suspended rope samples in the experiments
of Refs.~\cite{kociak,kasnew}.
An alternative way to estimate $N$ comes from atomic force microscopy,
which allows to measure the apparent radius of the rope, and hence yields
an estimate for the total number of tubes in the rope.  On average, 1/3
of the tubes are metallic \cite{nts},  and one should obtain
the same number $N$ from this approach. 
Fortunately, these two ways of estimating $N$ provide
consistent results in most samples \cite{kasnew}. Therefore the values
for $N$ used below are expected to be reliable.

Here we  always assume that phonon exchange leads to attractive 
interactions overcoming the (screened) Coulomb interactions.
This assumption can be problematic in ultrathin ropes, where practically
no screening arises unless there are close-by gate electrodes. 
For sufficiently large rope radius, however, theoretical arguments
supporting this scenario have been provided in Ref.~\cite{gonzalez3}.
In the absence of intra-tube disorder, then
the appropriate low-energy theory for  an
individual SWNT is the Luttinger liquid (LL) 
model \cite{egger97,kane97,ademarti}.
The LL theory of SWNTs 
is usually formulated within the Abelian  bosonization
approach \cite{gogolin}.  
With ${\bf x}=(x,\tau)$, where $x$ is the spatial 1D 
coordinate along the tube, and 
$0\leq \tau < 1/T$ is imaginary time, 
 and corresponding integration measure $d{\bf x}=dx d\tau$, 
the action for a single SWNT is 
\cite{egger97,kane97,ademarti}
\begin{eqnarray}\label{bosac}
S_{\rm LL} & = &  \int d{\bf x} \sum_{a=c\pm,s\pm}
 \frac{v_a}{2 g_a} \left[ (\partial_\tau \varphi_{a}/v_a)^2  + 
(\partial_x \varphi_{a})^2 \right] \nonumber \\ 
&=& \int d {\bf x } \sum_{a}
\frac{v_a g_a}{2} \left[ (\partial_\tau \theta_{a}/v_{a})^2  +  
(\partial_x \theta_{a})^2 \right],
\end{eqnarray}
which we take to be the same for every SWNT. 
Due to the electron spin and the additional K point degeneracy present
in nanotubes \cite{nts}, there are four channels, $a=c+,c-,s+,s-$,
corresponding to the total/relative charge/spin modes \cite{egger97,kane97},
with associated boson fields $\varphi_{a}({\bf x})$ and
dual fields $\theta_a({\bf x})$ \cite{gogolin}.
In the $a=(c+,s-)$ channels, the second (dual) formulation
turns out to be  more convenient, while the first line of Eq.~(\ref{bosac})
is more useful for $a=(s+,c-)$. 
The combined effect of Coulomb and phonon-mediated
electron-electron interactions results in the interaction parameter $g_{c+}$,
where we assume $g_{c+}>1$, reflecting effectively attractive
interactions \cite{ademarti}.  
In the neutral channels, there are only very weak
residual interactions, and we therefore put $g_{a\neq c+}=1$.
Finally, the velocities $v_a$ in Eq.~(\ref{bosac})
are defined as $v_a=v_F/g_a$, where
$v_F=8\times 10^5$~m$/$sec is the Fermi velocity.

Next we address the question which processes trigger the strongest
superconducting fluctuations in a nanotube rope.  This question
has been addressed in Refs.~\cite{gonzalez1,gonzalez2,ademarti}, 
and the conclusion of these studies is that Cooper pairs predominantly
form on individual SWNTs rather than involving electrons on different
SWNTs, see, e.g., the last section in Ref.~\cite{ademarti} for a 
detailed discussion.  Furthermore, the dominant intra-tube 
fluctuations involve {\sl singlet}\ (rather than triplet) Cooper 
pairs.  The relevant order parameter for superconductivity is then given by 
\cite{egger98}
\begin{equation}\label{orderpar2}
{\cal O}({\bf x})= \sum_{r\sigma\beta} \sigma \psi_{r,\sigma,\beta} 
({\bf x}) \psi_{-r,-\sigma,-\beta}({\bf x}) ,
\end{equation} 
where $\psi_{r\sigma \beta}$ denotes the electron field 
operator for a right- or left-moving electron ($r=\pm$) 
with spin $\sigma=\pm$ and K point degeneracy index $\beta=\pm$.
In bosonized language, this operator can be expressed as \cite{egger98}
\begin{eqnarray}\label{orderpar}
{\cal O}&=& \frac{1}{\pi a_0}
\cos[\sqrt{\pi} \theta_{c+}] \cos[\sqrt{\pi} \varphi_{c-}] \\
\nonumber &\times & 
\cos[\sqrt{\pi} \varphi_{s+}] \cos[\sqrt{\pi} \theta_{s-}] - 
(\cos\leftrightarrow \sin),
\end{eqnarray}
where we identify the  UV cutoff
necessary in the bosonization scheme with 
the graphite lattice constant, $a_0=0.24$~nm.
In what follows, we use the shorthand notation $\varphi_j$
to label all four boson fields $\varphi_a$ (or their dual fields)
corresponding to the $j$th SWNT, where $j=1,\ldots,N$.

The next step is to look at possible couplings among the
individual SWNTs.  In principle, three different processes
should be taken into
account, namely (i) direct Coulomb interactions,
(ii) Josephson couplings, and (iii) single-electron hopping.  
The last process is strongly
suppressed due to the generally different chirality of 
adjacent tubes \cite{kane}, and, in addition, for $g_{c+}>1$,
inter-SWNT Coulomb interactions are irrelevant \cite{schulz}.
Furthermore, as discussed in detail in Ref.~\cite{ademarti},
phonon-exchange mediated interactions between
{\sl different}\ SWNTs can always be neglected against
the intra-tube interactions.
Therefore the most relevant mechanism is Josephson
coupling between metallic SWNTs.  These couplings define
a Josephson matrix $\Lambda_{jk}$,
which contains the amplitudes for
Cooper pair hopping from the $j$th to the $k$th SWNT.
We put $\Lambda_{jj}=0$, and hence $\Lambda$ is a real, symmetric,
and traceless matrix.  It therefore has only real eigenvalues 
$\Lambda_\alpha$, which we take in descending order,
$\Lambda_1\geq \Lambda_2\geq \ldots\geq \Lambda_N$.
Moreover, there is at least one positive and at least one negative
eigenvalue.  The largest eigenvalue $\Lambda_1$ will be shown
to determine the mean-field critical temperature $T_c^0$ below.
The matrix $\Lambda$ is then expressed in the corresponding orthonormal
eigenbasis $|\alpha\rangle$,
\begin{equation}\label{lambdade}
\Lambda_{jk}= \sum_{\alpha} \langle j | \alpha \rangle \Lambda_\alpha
\langle \alpha | k \rangle ,
\end{equation}
where $\langle j | \alpha \rangle$ is the  
real orthogonal transformation from the basis of lattice 
points $\{|j\rangle\}$ to the basis
$\{|\alpha\rangle\}$ that diagonalizes $\Lambda$.
Clearly,  $\langle j | \alpha \rangle = \langle \alpha | j \rangle $.  
In what follows, we define $\alpha_0$ 
such that $\Lambda_{\alpha}>0$ for $\alpha<\alpha_0$.

The Euclidean action of the rope is then
\begin{equation}\label{ea}
S = \sum_{j=1}^N S_{\rm LL}[\varphi_{j}] - \sum_{jk} 
\Lambda_{jk}\int d{\bf x} \, {\cal O}^\ast_j {\cal O}_k^{},
\end{equation}
where ${\cal O}_j$ is the order parameter specified in Eq.~(\ref{orderpar}).
The action (\ref{ea}) defines the model that is 
studied in the remainder of our paper.  
For studies of closely related models, see
also Refs.~\cite{schulz,carr}. 

In order to decouple the Josephson term in Eq.~(\ref{ea}),  
we employ a Hubbard-Stratonovich transformation. 
To that purpose,
since the Josephson matrix has at least one negative eigenvalue,
we first express $\Lambda$ in its eigenbasis, see Eq.~(\ref{lambdade}).
The Josephson term in Eq.~(\ref{bosac}) is then rewritten as
\[
\sum_{jk}
{\cal O}_j^* \Lambda_{jk} {\cal O}^{}_k = \sum_{\alpha}\ 
{\rm sgn}(\Lambda_\alpha) 
| \Lambda_\alpha |  {\cal O}_\alpha^* {\cal O}^{}_{\alpha} ,
\]
where the order parameter in the $|\alpha\rangle$ basis is 
\begin{equation}\label{expa00}
{\cal O}_\alpha^{} \equiv \sum_i \langle \alpha | i 
\rangle {\cal O}^{}_i, \quad
{\cal O}_\alpha^* \equiv \sum_i {\cal O}_i^* \langle i | \alpha \rangle.
\end{equation}
By introducing a field $\Delta^{}_\alpha({\bf x})$ for each 
Josephson eigenmode \cite{foot1},
with (formally independent) complex conjugate field
$\Delta^\ast_\alpha$, it is now possible to 
perform the Hubbard-Stratonovich transformation following
the standard procedure \cite{nagaosa1}.
With integration measure ${\cal D} \Delta  = \prod_{\alpha} 
{\cal D}\Delta^*_\alpha  
{\cal D}\Delta^{}_\alpha$,  the effective action entering
the partition function $Z=\int {\cal D}\Delta \exp(-S_{\rm eff}[\Delta])$ 
reads
\begin{equation} 
\label{effac}
S_{\rm eff}[\Delta]= S_0[\Delta] + \int d{\bf x} \sum_{\alpha} 
 \Delta^\ast_\alpha  \frac{1}{|\Lambda_\alpha|} 
\Delta_\alpha ,
\end{equation}
where the action $S_0[\Delta]$ is formally defined via the remaining 
path integral over the boson fields $\varphi_{j}$,
\begin{eqnarray} \label{f00}
S_0[\Delta] &=& -\ln \int
\prod_{j=1}^N {\cal D}\varphi_{j} \, e^{ -\sum_j S_{\rm LL}[\varphi_{j}]} 
\times \nonumber \\
& \times & e^{- \int d{\bf x}
\sum_{\alpha} c_{\alpha} \left(
\Delta^*_\alpha {\cal O}^{}_\alpha +
{\cal O}^*_\alpha \Delta^{}_\alpha
\right)}, 
\end{eqnarray}
with $c_\alpha = 1$ for $\alpha <\alpha_0$,
and $c_\alpha = i$ otherwise.

\section{Quantum Ginzburg-Landau approach}\label{sec3}

\subsection{Cumulant expansion}

Clearly, closed analytical evaluation of the path integral
in Eq.~(\ref{f00}) is in general impossible. 
In order to make progress, approximations are necessary,
and in the following we shall construct and analyze the
 Ginzburg-Landau (GL) action 
\cite{nagaosa1,tinkham} for this problem.
It turns out to be essential to take into account
quantum fluctuations, i.e., the imaginary-time dependence of the
order parameter field $\Delta_\alpha(x,\tau)$.
In the standard (static) Ginzburg-Landau theory, such effects are
ignored.   

The derivation of the GL action
proceeds from a cumulant expansion 
of Eq.~(\ref{f00}) up to quartic order in the $\Delta_\alpha$.
This is a systematic expansion 
in the parameter $|\Delta|/2\pi T$ \cite{nagaosa1}, 
and by self-consistently computing this parameter, one
can determine the regime of validity of GL theory.
We stress that this expansion is {\sl not}\ restricted to $N\gg 1$.
In addition, for the long-wavelength low-energy regime of primary interest
here, we are entitled to perform a gradient
expansion.  
Using the single-chain correlation function
$G({\bf x}_{12})=\langle {\cal O}({\bf x}_1)
{\cal O}^\ast({\bf x}_2)\rangle$ 
of the operator ${\cal O}$ in Eq.~(\ref{orderpar})
 with respect to the free boson action
$S_{\rm LL}$,  and the connected four-point correlation function
\begin{eqnarray*}
G^{(4)}_c({\bf x}_1,{\bf x}_2,{\bf x}_3,{\bf x}_4) &=& 
\langle {\cal O}({\bf x}_1) 
{\cal O}({\bf x}_2) {\cal O}^*({\bf x}_3) {\cal O}^*({\bf x}_4)\rangle 
 \\
&-& \langle {\cal O}({\bf x}_1) {\cal O}^*({\bf x}_3) \rangle
\langle {\cal O}({\bf x}_2) {\cal O}^*({\bf x}_4)\rangle \\  
&-&\langle {\cal O}({\bf x}_1) {\cal O}^*({\bf x}_4) \rangle
\langle {\cal O}({\bf x}_2) {\cal O}^*({\bf x}_3)\rangle ,
\end{eqnarray*}
the cumulant-plus-gradient expansion 
up to quartic order yields for
the effective Lagrangian density 
\begin{eqnarray}\label{lagal}
L[\Delta] &=& \sum_{\alpha<\alpha_0}
\Bigl[ C\, |\partial_x \Delta^{}_\alpha|^2 + 
D\, |\partial_\tau \Delta^{}_\alpha |^2 \\ 
&+&  \left( \Lambda_\alpha^{-1} - A \right) |\Delta^{}_\alpha|^2 
\Bigr]  \nonumber \\ \nonumber
&+& B \sum_{\alpha_i<\alpha_0} 
f^{\alpha_1,\alpha_2}_{\alpha_3,\alpha_4}
\, \Delta^*_{\alpha_1}
\Delta^*_{\alpha_2} \Delta^{}_{\alpha_3} 
\Delta^{}_{\alpha_4} ,
\end{eqnarray}
where we use the notation
\[
f^{\alpha_1,\alpha_2}_{\alpha_3,\alpha_4}=\sum_i 
\langle \alpha_1 | i\rangle \langle \alpha_2 | i\rangle
\langle i | \alpha_3 \rangle
\langle i | \alpha_4 \rangle .
\]
The temperature-dependent positive coefficients $A, B, C, D$ are obtained as
\begin{eqnarray} 
A  &=& \int d{\bf x} \, G({\bf x}) ,\label{coefA}\\
\label{coefB}
B & = & -\frac{1}{4}\int d{\bf x}_{1} d{\bf x}_2 
d{\bf x}_3 \, G^{(4)}_c({\bf x}_1,{\bf x}_2,{\bf x}_3,{\bf x}_4), \\
C &=& \frac12 \int d{\bf x} \, x^2 G({\bf x}) ,\label{coefC}\\
D &=& \frac12 \int d{\bf x} \, \tau^2 G({\bf x}) .\label{coefD}
\end{eqnarray}
Due to translation invariance, the integral for $B$
does not depend on ${\bf x}_4$.
Besides temperature, these coefficients basically depend only 
on the important LL interaction parameter $g_{c+}$.
In particular, as it is discussed below, for $g_{c+} >1$, 
the coefficient $A$ grows as $T$ is lowered. 
For static and uniform configurations, modes with
$\alpha>\alpha_0$ never become critical.  
One can then safely integrate over these modes, which leads to a 
renormalization of the parameters governing the remaining 
modes. Such renormalization effects are however tiny,
and  thus are completely neglected in Eq.~(\ref{lagal}).

At this stage, it is useful to switch to
an order parameter field defined on the $j$th SWNT,
\begin{equation}\label{opontube}
\Delta_j=\sum_{\alpha<\alpha_0} \langle j| \alpha \rangle \Delta_\alpha .
\end{equation}  
After some algebra, the Lagrangian density (\ref{lagal}) can 
be written as
\begin{eqnarray}
L[\Delta]  =  \sum_{j=1}^N \Bigl[
C  |\partial_x \Delta_j^{}|^2 + 
D  |\partial_\tau \Delta_j^{}|^2 + B |\Delta_j|^4 &+& \nonumber \\
+ \left( \Lambda_1^{-1} - A \right) |\Delta^{}_j|^2 \Bigr] 
+ \sum_{jk}\Delta^\ast_j V_{jk} \Delta^{}_k , &{}&  
\label{hs2} 
\end{eqnarray}
with the real, symmetric, and positive definite matrix
\begin{equation}\label{vjk}
V_{jk}=\sum_{\alpha< \alpha_0} 
\langle j | \alpha \rangle  \left( \Lambda_\alpha^{-1} -
\Lambda_1^{-1} \right) \langle \alpha | k \rangle. 
\end{equation}
Notice that, strictly speaking,
 the fields $\Delta_i$ are not all independent, 
because we have defined them from the subset of positive modes.
The transformation in Eq.~(\ref{opontube}) is indeed not invertible.  
Nevertheless, in the following, we treat them as formally independent.
This only affects the precise values of the $V_{ij}$ but does
not qualitatively change our results.
The expectation value of the order parameter field
(\ref{orderpar}) can be expressed in terms of linear combinations
of the fields $\Delta_j(x,\tau)$, and hence it is indeed justified
to call $\Delta_j$ a  proper ``order parameter field''.

Equation (\ref{hs2}) specifies the full GL action, taking
into account quantum fluctuations and transverse modes
for arbitrary number $N$ of active SWNTs. 
In the limit $N\to\infty$, and considering only static field
configurations, results similar to those of Ref.~\cite{schulz} are
recovered.  In that limit the last term in Eq.~(\ref{hs2}) gives 
indeed the gradient term in the transverse direction, and one
obtains the standard 3D GL Lagrangian.
There is however an important difference, namely the starting point of 
Ref.~\cite{schulz} is a model of Josephson-coupled 
1D superconductors, whereas we start from an array
of metallic chains with $g_{c+} > 1$, where the 
inter-chain Josephson coupling is crucial in stabilizing
superconductivity.
More similar to ours is the model investigated in Ref.~\cite{carr}.
However, in that paper, the metallic chains are assumed to have a spin gap,
which is not the case for the SWNTs in a rope in the 
temperature range of interest. Furthermore, the main focus 
in Ref.~\cite{carr} is the competition between charge density wave and
superconducting instabilities, whereas in our case, as discussed above,
the formation of a charge density wave is strongly suppressed
by compositional disorder, i.e., different chiralities of adjacent 
tubes, and we do not have to take the corresponding instability 
into account.

\subsection{Ginzburg-Landau coefficients}

In order to make quantitative predictions, it is necessary to compute
the GL coefficients defined in Eqs.(\ref{coefA})-(\ref{coefD}).   
While this is possible in principle for the
full four-channel model (\ref{bosac}), here we will instead
derive the coefficients
for a simpler model, where the K point degeneracy is neglected.
This leads to an effective spin-$1/2$ Luttinger liquid action
with interaction parameter $g_c$ ($g_s=1$) 
and velocity $v_c=v_F/g_c$ ($v_s=v_F$).
Up to a prefactor of order unity, the respective results can be matched onto
each other.  This can be made explicit, e.g., for the coefficient $A$,
where we get from the full action (\ref{bosac}) 
\[
A= \frac{c}{v_F}
\left( \frac{\pi a_0 T}{v_{c+}} \right)^{(g_{c+}^{-1}-1)/2} .
\]
The proportionality constant $c$ is 
found to differ from $\tilde A/2\pi^2$ 
[see Eq.~(\ref{coefA2}) below, which follows from
the spin-$1/2$ description] only by a factor of order unity. 
In the simpler model neglecting the K point
degeneracy, one then needs to take 
\[
g_c^{-1}=\frac{1+g^{-1}_{c+}}{2},
\]
which gives, for $g_{c+}=1.3$, a value of $g_c \approx 1.1$.
This way, all exponents of the resulting power-law correlation functions
(which are the physically relevant quantities)
in the ``reduced'' model 
are the same as in the complete model, and only 
prefactors of order unity may be different 
for the respective GL coefficients.
The bosonized order parameter (\ref{orderpar}) in the
simpler model is then given by
\[
{\cal O}= \frac{1}{\pi a_0} \cos[\sqrt{2\pi} \varphi_s]
\exp[i\sqrt{2\pi}\theta_c].
\]
Using the finite-temperature correlation functions of 
the fields $\theta_c$ and $\varphi_s$ \cite{gogolin},
\begin{eqnarray*}
\langle \theta_c({\bf x})\theta_c({\bf 0})\rangle &=& \frac{-1}{2\pi g_c} 
\ln \left(\frac{v_c}{\pi a_0 T} 
\left |\sinh \frac{\pi T (x+iv_c\tau)}{v_c}\right|\right) , \\
\langle \varphi_s({\bf x}) \varphi_s({\bf 0}) \rangle &=& \frac{-1}{2\pi} 
\ln \left(\frac{v_F}{\pi a_0 T}
\left|\sinh \frac{\pi T(x+iv_F\tau)}{v_F}\right|\right), 
\end{eqnarray*}
and rescaling the integration variables $x$ and $\tau$ 
in Eqs.~(\ref{coefA})-(\ref{coefD}), 
 explicit expressions follow in the form
\begin{eqnarray}
A(T) &=& \frac{1}{2\pi^2 v_F}  ( \pi a_0 T/v_c )^{g_c^{-1}-1}
\tilde{A},\label{coefA2}\\ \nonumber
B(T) &=& \frac{a_0^2 }{32\pi^4 v_c v_F^2} 
 ( \pi a_0 T/v_c )^{2g_c^{-1} -4} \tilde{B}, \\ \nonumber
C(T) &=& \frac{a_0^2}{4\pi^2 v_F} 
 ( \pi a_0 T/v_c )^{g_c^{-1}-3} \tilde{C},\\
D(T) &=& \frac{a_0^2}{4\pi^2 v_Fv_c^2} 
 ( \pi a_0 T/v_c )^{g_c^{-1}-3} \tilde{D}. \nonumber
\end{eqnarray}
Dimensionless $g_c$-dependent numbers 
$\tilde A, \tilde B, \tilde C, \tilde D$
were defined as follows.  With the notation ${\bf z}=(w,u)$
and 
\[
\int d{\bf z}= 
 \int_{0}^{\pi} du \int_{-\infty}^\infty dw,
\]
we have
\begin{eqnarray*}
\tilde{A}&=& \int \frac{d{\bf z}}{ f_c({\bf z}) f_s({\bf z})}, \\
\tilde{C}&=& \int d{\bf z} \frac{w^2}{ f_c({\bf z}) f_s({\bf z})},\\ 
\tilde{D}&=& \int d{\bf z} \frac{u^2}{ f_c({\bf z}) f_s({\bf z})}, 
\end{eqnarray*}
where functions $f_{c,s}$ are introduced as
\begin{eqnarray*}
f_c({\bf z}) &=& |\sinh(w+iu)|^{1/g_c},\\
f_s({\bf z})  &=& |\sinh(w/g_c + iu)|.
\end{eqnarray*}
The coefficient of the quartic term in the GL functional is
\begin{eqnarray*}
& &\tilde B = \int \frac{d{\bf z}_1 d{\bf z}_2 d{\bf z}_3}
{f_c({\bf z}_2) f_c({\bf z}_{13})} \Biggl[   
\frac{4}{f_s({\bf z}_2)f_s({\bf z}_{13})}-
 \frac{f_c({\bf z}_1) f_c({\bf z}_{23})}
    {f_c({\bf z}_3)f_c({\bf z}_{12})} \\ &\times&  \left( 
   \frac{f_s({\bf z}_1)f_s({\bf z}_{23})} 
{f_s({\bf z}_2)f_s( {\bf z}_{13}) f_s({\bf z}_3)f_s({\bf z}_{12})}
+ (1\leftrightarrow 2) + (1 \leftrightarrow 3) \right)  
\Biggr ] 
\end{eqnarray*}
with ${\bf z}_{ij}=(w_i-w_j, u_i-u_j)$. 
 The quantity $\tilde{B}$
is evaluated using the Monte Carlo method.
For $g_c=1$, we first numerically reproduced
the exact result $\tilde{B}=8\pi^2 \tilde{C}$ 
with $\tilde{C}=7\pi \zeta(3)/4$ \cite{nagaosa1}.
Numerical values can then be obtained for arbitrary $g_c$.
Numerical evaluation yields for $g_c \approx 1.1$ (corresponding to
$g_{c+}=1.3$) the following results:
\begin{equation}\label{tildes}
\tilde{A} \simeq 17.4 , \quad \tilde{B} \simeq 392(1) , 
\quad \tilde{C}\simeq 8.15 ,\quad \tilde{D}\simeq 6.97 . 
\end{equation}

\subsection{Mean-field transition temperature}

Since in the rope only a modest number of transverse modes are present, 
a natural definition of the mean-field critical temperature $T_c^0$
is the temperature at which the mode corresponding to 
the largest eigenvalue of $\Lambda$ 
becomes critical. {}From Eq.~(\ref{hs2}), this leads to 
the condition $A(T)=\Lambda_1^{-1}$,  and hence to
 the mean-field critical temperature
\begin{equation}\label{tc}
T_c^0 = \frac{v_c}{\pi a_0} \left( \frac{\tilde A \Lambda_1}{2\pi^2 v_F}
\right)^{g_c/(g_c-1)} ,
\end{equation}
which exhibits a dependence on the number $N$ of 
active SWNTs in the rope through $\Lambda_1$.  For large
$N$, the eigenvalue $\Lambda_1$ saturates, and Eq.~(\ref{tc})
approaches the bulk transition temperature. 

To provide concrete theoretical
 predictions for $T_c^0$ is difficult,
since the Josephson matrix is in general unknown, and the 
results for $T_c^0$ very sensitively depend on $\Lambda_1$.
Using estimates of Ref.~\cite{gonzalez2} and typical $N$ as reported
in Ref.~\cite{kasnew},  as an order-of-magnitude estimate,
we find  $T_c^0$ values around $0.1$ to 1~K.  When comparing 
to experimental results, $\Lambda_1$ can be inferred from the
actually measured $T_c$, which in turn provides values in 
reasonable agreement with theoretical expectations \cite{gonzalez1}.

\subsection{Low-energy theory: $T < T_c^0$}

\begin{figure}
\centerline{\epsfxsize=7cm\epsfysize=6cm 
\epsffile{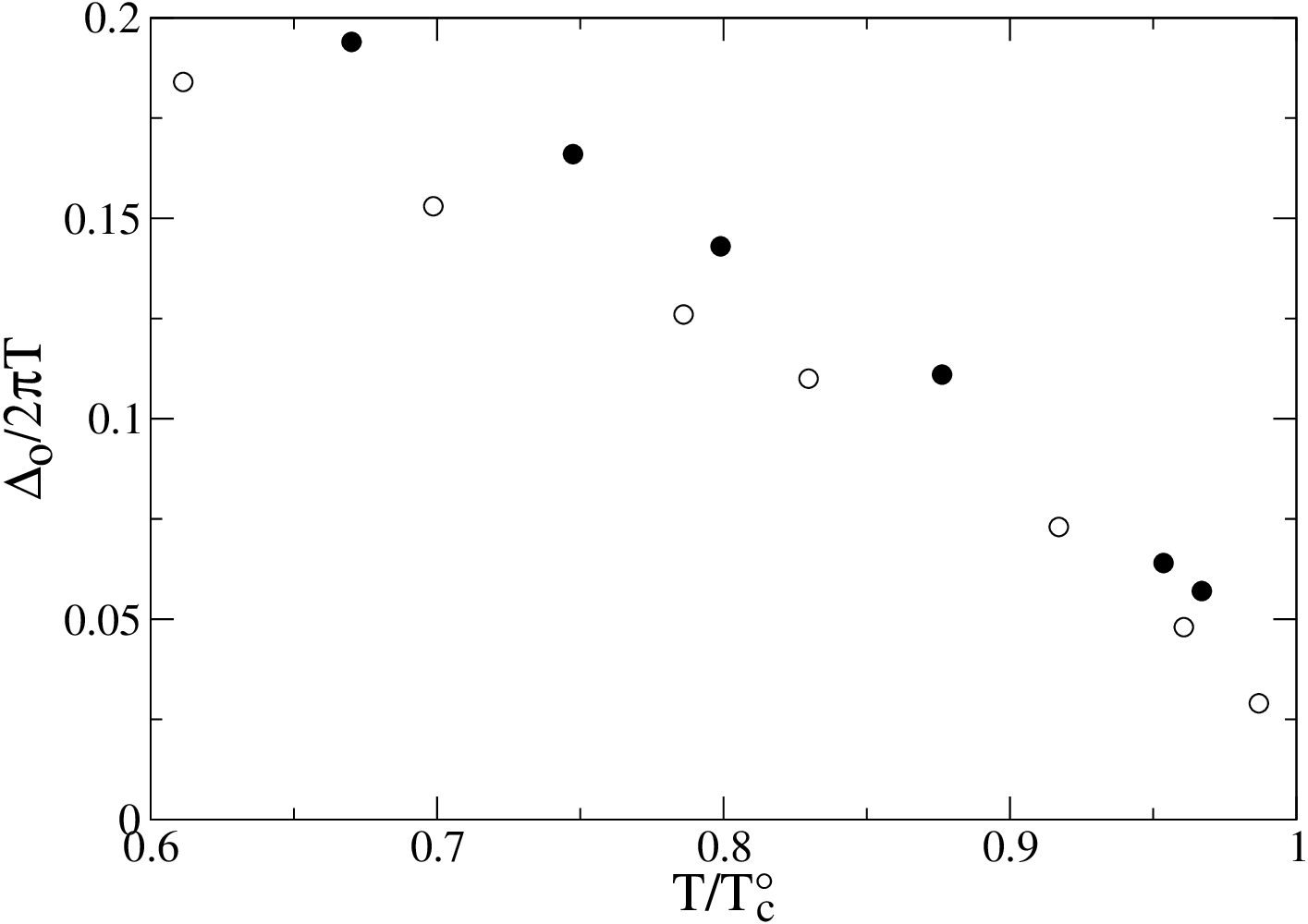}}
\caption{\label{fig1}  
Temperature dependence of $\Delta_0/2\pi T$ versus $T/T_c^0$ for
$N=31$ (open circles) and $N=253$ (filled circles).  }
\end{figure}

In what follows, we focus on temperatures $T<T^0_c$.  
Then it is useful to employ an amplitude-phase representation of the
order parameter field,
\begin{equation}\label{densphas}
\Delta_j ({\bf x})=|\Delta_j|({\bf x}) \exp [i\phi_j({\bf x})],
\end{equation}
where the amplitudes $|\Delta_j|$ are expected to be
finite with a gap for fluctuations
around their mean-field value.  At not too low temperatures, the GL action
corresponding to Eq.~(\ref{hs2}) is accurate (see below), 
and the mean-field values
follow from the saddle-point equations.
Considering only static and uniform field configurations, 
we find $\phi_i\equiv \phi$, where in principle also other (frustrated)
configurations with $\phi_i-\phi_j=\pm \pi$ could contribute.
Such configurations presumably correspond to maxima of the free energy,
and are ignored henceforth.
The saddle-point equations then reduce to equations for the amplitudes alone,
\begin{equation}\label{gap}
\sum_{j} V_{ij} |\Delta_j| + (\Lambda_1^{-1}-A)|\Delta_i| +
 2B |\Delta_i|^3 =0 ,
\end{equation}
whose solution yields the transverse order parameter profile.
Numerical study of Eq.~(\ref{gap}) using a standard Newton-Raphson
root-finding algorithm then allows to 
extract the profile $\{|\Delta_j|\}$ 
for a given Josephson matrix $\Lambda_{ij}$.
We briefly discuss the solution of Eq.~(\ref{gap})
for the idealized model of a rope as a trigonal 
lattice exclusively composed of
$N$ metallic SWNTs, where $\Lambda_{ij}=\lambda$ for nearest
neighbors $(i,j)$, and $\Lambda_{ij}=0$ otherwise.  For this model,
Fig.~\ref{fig1} shows the  resulting average amplitude 
$\Delta_0=\sum_i |\Delta_i|/N$ as a function of temperature for
$\lambda/v_F=0.1$ and two values of $N$.
Since $\Delta_0/2\pi T$ is the expansion parameter entering
the construction of the GL functional, and it
remains small down to $T\approx T_c^0/2$, we
conclude that the GL theory is self-consistently valid 
in a quantitative way down to such temperature
scales. In our discussion below, GL theory turns 
out to be qualitatively useful even down to $T=0$. 

Fixing the amplitudes $|\Delta_j|$ at their mean-field values, and
neglecting the massive amplitude fluctuations around these values, 
the Lagrangian follows from Eq.~(\ref{hs2}) as
\begin{eqnarray}\label{final}
L&=& \sum_{j=1}^N
 \frac{\mu_j}{2\pi} \left[ c_s^{} (\partial_x\phi_j)^2 + c_s^{-1}
(\partial_\tau \phi_j)^2 \right]\\ \nonumber
& + & \sum_{i>j} 2V_{ij}|\Delta_i||\Delta_j| 
\cos(\phi_i-\phi_j) ,
\end{eqnarray}
with the Mooij-Sch\"on velocity \cite{mooij},
\begin{equation}\label{css}
c_s \equiv v_c \sqrt{\tilde{C}/\tilde{D}}, 
\end{equation} 
and dimensionless phase stiffness parameters
\begin{equation}\label{stiff1}
\mu_j= 2\pi C |\Delta_j|^2 /c_s.
\end{equation}
At this stage, electromagnetic potentials can be coupled in via standard
Peierls substitution rule \cite{nagaosa1}, 
and dissipative effects due to the electromagnetic environment
can be incorporated following Ref.~\cite{blatter}.

\section{1D action and quantum phase slips}\label{sec4}

\subsection{1D phase action}\label{sec41}

Numerical evaluation of Eq.~(\ref{gap}) 
shows that for $T$ well below $T_c^0$,
transverse fluctuations are heavily suppressed.
While this statement only applies to amplitude fluctuations,
one can argue that also the transverse phase fluctuations 
are strongly suppressed.  The basic argument relates to 
the scaling dimension [in the renormalization group (RG) sense]
of the operator $\cos(\phi_i-\phi_j)$, which is essentially
governed by the $\mu_j$. For $T$ well below $T_c^0$, the 
$\mu_j$ become large, and the cosine operators get strongly
relevant, locking the phases all together.  In the low-temperature
regime of main interest below, this argument allows to substantially
simplify Eq.~(\ref{final}).
Then also no detailed knowledge about the Josephson 
matrix is required, because the only relevant information 
is essentially contained in $T^0_c$.

Putting all phases $\phi_j=\phi$, 
we arrive at a standard (Gaussian) 1D superconducting phase action 
\cite{tinkham},
\begin{equation}\label{finala}
S=\frac{\mu}{2\pi}\int dx d\tau \left[
c_s^{-1}(\partial_\tau\phi)^2 + c_s (\partial_x \phi)^2 \right],
\end{equation} 
with dimensionless rigidity
$\mu = \sum_j \mu_j$, see Eq.~(\ref{stiff1}), 
and $c_s$ as given in Eq.~(\ref{css}).
Assuming GL theory to work even down to $T=0$ for the moment, and neglecting 
the $V_{ij}$-term in Eq.~(\ref{gap}), a simple analytical estimate follows
in the form
\begin{equation} \label{mu1}
\mu(T) = N \nu \left[ 1 - (T/T_c^0)^{(g_c-1)/g_c}\right] , 
\end{equation}
where the number $\nu$  is
\begin{equation} \label{defnu}
\nu =4\pi \tilde{A} (\tilde{C}\tilde{D})^{1/2}  / \tilde{B}.
\end{equation}
The peculiar temperature dependence of the phase stiffness in Eq.~(\ref{mu1}),
reflecting the underlying LL physics of the individual SWNTs, is one of 
the main results of this paper.
In the effective spin-$1/2$ description employed here,
using the numbers specified
in Eq.~(\ref{tildes}) for $g_c=1.1$ results in $\nu\approx 4$.
Remarkably, at
$T=0$,  Eq.~(\ref{mu1}) coincides, up to a prefactor of order unity,
with the rigidity $\bar{\mu}$ obtained from standard 
mean-field relations \cite{nagaosa1},
\[
\bar{\mu}= \pi^2 n_s R^2/2 m^\ast c_s = \bar{\nu} N.
\]
With the density of condensed electrons $n_s$ and rope radius $R$, 
this implies $\bar{\nu}\approx v_F/c_s$, which is of order unity. 
We therefore conclude that the GL prediction (\ref{mu1}) for $\mu(T)$ is 
robust and useful even outside its strict validity regime.

The result (\ref{mu1}) for the stiffness is central for the
following discussion. The value we obtain for $\nu$, however,
should not be taken as a very precise estimate.
First, it can be affected
by factors of order unity under a full four-channel calculation 
taking into account the K point degeneracy,
as this affects each of the numbers in Eq.~(\ref{tildes}) by a 
factor of order unity.  Second, uncertainties in the parameter $g_c$ 
will also affect $\nu$ by a factor of order unity.  Moreover, based
on the discussion in Ref.~\cite{zaikin2}, one expects
on general grounds that intra-SWNT disorder and
dissipative effects, both of which are not included 
in our model, will  effectively lead to a {\sl decrease} of the parameter
$\nu$ entering  Eq.~(\ref{mu1}).
Therefore $\nu$ is taken below as a fit
parameter when comparing to experimental data. 
Since the number of active SWNTs $N$ can be estimated 
from the residual resistance, 
and the transition temperature $T_c$, see Eq.~(\ref{tc1}) below, can be
determined  from the experimentally observed transition temperature, 
 $\nu$ is basically the only free remaining parameter. 
Fits of our theoretical results 
to experimental data are then expected to yield values for $\nu$ around 
$\nu\approx 1$. This is verified below in Sec.~\ref{sec6}.

\subsection{Phase slips}

In the 1D situation encountered here, superconductivity
can be destroyed by phase slips \cite{tinkham}. 
A phase slip (PS) can be visualized as a process in which
fluctuations locally destroy the amplitude of the superconducting 
order parameter, which effectively disconnects the 1D
superconductor into two parts. Simultaneously, the phase, 
being defined only up to $2\pi$, is allowed to ``slip'' by $2\pi$
across the region where the amplitude vanishes. 
This process then leads to finite dissipation in the superconducting wire
via the
Josephson effect. Depending on temperature, phase slips 
can be produced either by thermal or by quantum fluctuations. 
In the first case, which is commonly realized very near the critical 
temperature, we have a thermally activated 
phase slip (TAPS). At lower temperature, 
the quantum tunneling mechanism dominates, 
and one speaks of a quantum phase slip (QPS).
For a textbook description of quantum phase slips, see Ref.~\cite{chaikin}. 
Below we demonstrate that in superconducting ropes, only QPSs
are expected to play a prominent role. 

A QPS is a topological vortex-like excitation of the
superconducting phase field $\phi(x,\tau)$
that solves the equation of motion for
the action (\ref{finala}) with a 
singularity at the core, where superconducting order 
is locally destroyed and a phase cannot be defined.
Defining a thermal lengthscale as
\begin{equation}\label{lt}
L_T=  c_s/ \pi T  ,
\end{equation}
for rope length $L\to \infty$ and $L_T\to \infty$,
a QPS with core at $(x_i,\tau_i)$ and winding number $k_i=\pm 1$ 
(higher winding numbers are irrelevant)
is given by \cite{chaikin}
\begin{equation}\label{qps}
\phi(x,\tau)=k_i \arctan 
\left[\frac{c_s(\tau-\tau_i)}{(x-x_i)}\right] ,
\end{equation}
where the finite $L,L_T$ solution 
follows by conformal transformation \cite{blatter}.
The action of a QPS consists of two terms, one associated with the local
loss of condensation energy, the core action $S_c$, and  
the other with the vortex strain energy. % $S_{\rm el}$.
While a detailed computation of $S_c$ requires a microscopic 
description of the dynamics inside the vortex core \cite{zaikin2}, 
a simple qualitative argument is able to predict an order-of-magnitude
estimate $S_c\approx \mu/2$ \cite{chaikin}.

This result allows us to assess the relative contribution
of the TAPS and QPS mechanisms.  
The production rate for the creation of one
vortex is \cite{zaikin2} 
$\gamma_{\rm QPS}\approx \frac{S_c L c_s}{\kappa} \exp(-S_c)$,
where $\kappa$ is the core size.
Within exponential accuracy, comparing this formula to the respective 
standard TAPS rate expression \cite{tinkham},  
the crossover temperature from
TAPS- to QPS-dominated behavior is 
$T^*_{\rm PS}= 2\Delta F/ N\nu$,
with activation barrier $\Delta F$.  Using results of 
Ref.~\cite{carr}, we estimate the latter as 
$\Delta F=8\sqrt{2} R(g_c) N T_c^0/3$,
with dimensionless coefficient $R(g_c)$ of order unity.  
Finally, this implies $T^*_{\rm PS}\approx  T_c^0$.
Since the true transition temperature $T_c<T_c^0$, see below,
in the temperature regime $T<T_c$, the influence of a TAPS 
can safely be neglected against the QPS.

The generalization to many QPSs then leads to the standard picture 
of a Coulomb gas of charges $k_i=\pm 1$, with fugacity $y=e^{-S_c}$, total 
charge zero, and logarithmic interactions \cite{nagaosa1,chaikin}.  
The partition function  $Z=Z_G Z_V$ contains a regular factor 
$Z_G$ and the vortex contribution
\begin{equation} \label{partfncvor}
Z_V = \sum_{n=0}^\infty \frac{y^{2n}}{(n!)^2} 
 \int \frac{\prod_{m=1}^{2n} d{\bf r}_m}{(c_s\kappa^2)^{2n}}
\sum_{\{k \}} e^{\mu \sum_{i\neq j} k_i k_j \ln (r_{ij}/\kappa)}. 
\end{equation}
This model undergoes a Berezinski-Kosterlitz-Thouless
transition driven by the nucleation of vortices, here corresponding to
a transition from a phase $\mu>\mu^*$,  where QPSs are 
confined into  neutral pairs and the rope forms a 1D superconductor 
with finite phase stiffness 
and quasi-long-range order,  to a phase $\mu <\mu^*$ where
QPSs proliferate.  In that phase, vortices  are deconfined and  
destroy the phase stiffness, thereby producing normal behavior,
where ``normal'' does of course not imply Fermi-liquid behavior.
The phase boundary is located at $\mu^*=2+4\pi y \simeq 2$. 
The true transition temperature $T_c$ is therefore {\sl not}\ the mean-field
transition temperature $T_c^0$, but follows
from the condition $\mu(T_c)=\mu^\ast$. Putting $\mu^\ast=2$, 
Eq.~(\ref{mu1}) yields
\begin{equation}\label{tc1}
T_c/T_c^0 = \left[1- 2/ N\nu \right]^{g_c/(g_c-1)}.
\end{equation}
This $T_c$ depression is quite sizeable for $N\alt 100$. To give 
concrete numbers, taking $\nu=1$, for $N=25,50,$ and 100, the 
ratio $T_c/T_c^0$ equals $0.40, 0.63,$ and $0.80$, respectively.
QPSs also have an important and observable effect in the 
superconducting regime, as will be discussed in the next section.

\begin{figure}
\centerline{\epsfxsize=7cm \epsfysize=6cm
\epsffile{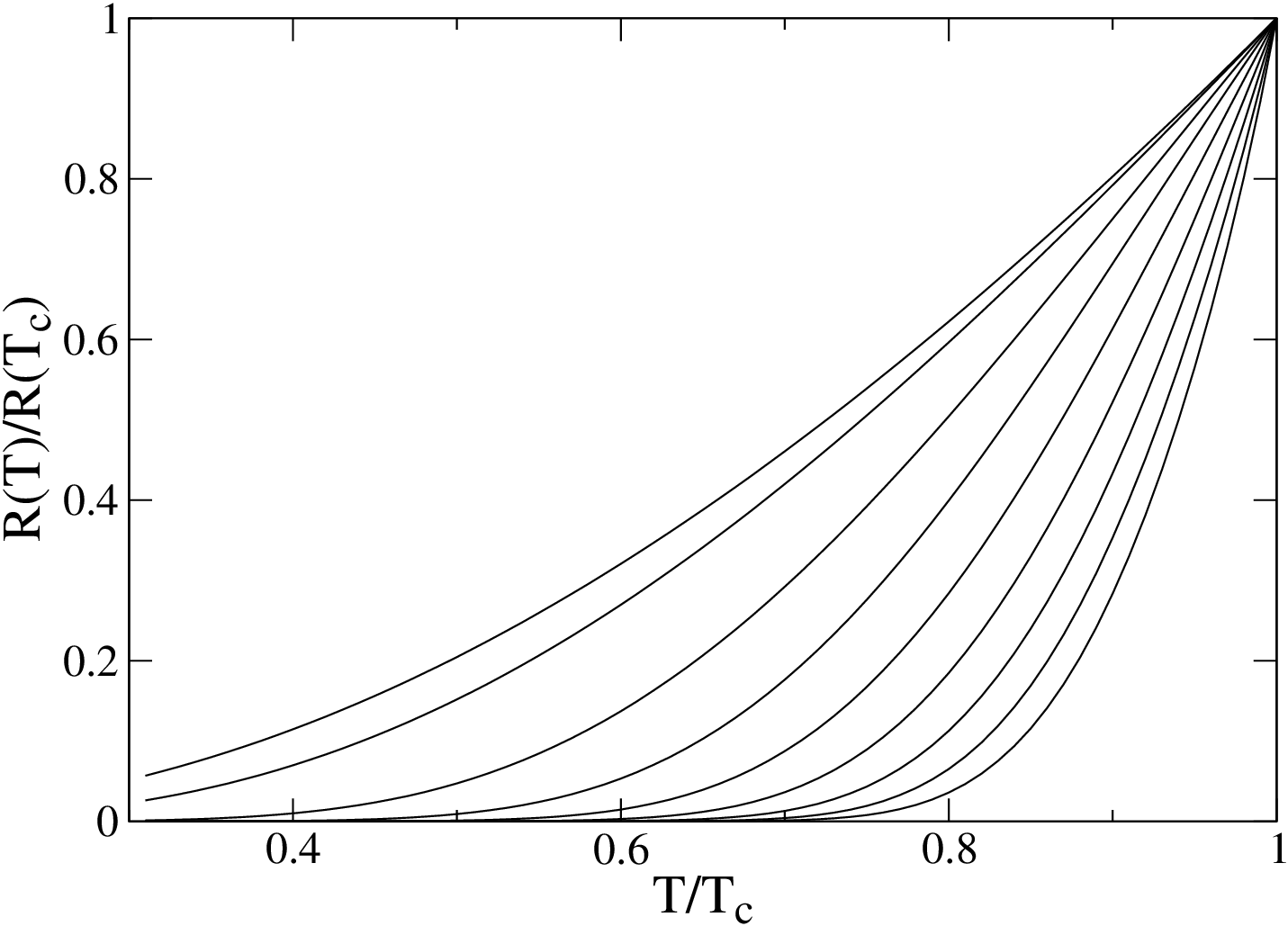}}
\caption{\label{figevo}  
Temperature-dependent resistance $R(T<T_c)$ 
predicted by Eq.~(\ref{resis})
for $\nu=1$ and different $N$. The smaller is $N$, the broader is
the transition. From the leftmost to the rightmost curve, 
$N=4,7,19,37,61,91,127,169,217$.
}
\end{figure}

\section{Resistance below $T_c$}\label{sec5}

A phase slip produces finite dissipation through the Josephson
effect, and therefore introduces a finite resistance even
in the superconducting state, $T<T_c$.   The QPS-induced linear
resistance $R(T)=V/I$ for $T<T_c$ can be computed perturbatively in the
QPS fugacity $y$ \cite{zaikin1}. For that purpose, we imagine that one 
imposes a small current $I$ to flow through the rope.
The presence of QPSs implies that a voltage drop $V$ occurs,
which is related to the average change in phase,
\[
V=\frac{\langle \dot\phi \rangle}{2e}=
\frac{\pi}{e}[\Gamma(I)-\Gamma(-I)] ,
\]
where $\Gamma(\pm I)$  is the rate for a phase slip 
by $\pm 2\pi$ \cite{zaikin1}.  This rate can be obtained
following Langer \cite{langer} as
the imaginary part acquired by the 
free energy $F(I)$ under an appropriate analytic continuation,
\begin{equation}
\Gamma(\pm I)= -2 \, {\rm Im} \, F(\pm I) . 
\end{equation} 
We  only consider the contribution of a single pair of QPSs, 
i.e., compute $R(T)$ to second order in $y$.  
Expanding Eq.~(\ref{partfncvor}) to order $y^2$, 
the free energy at this order reads
\begin{equation}
F = - \frac{Ly^2 c_s^2}{\kappa^4} \int_0^{1/T} d\tau 
\int_{-L/2}^{L/2} dx \, e^{\epsilon \tau -2 \mu g_E(x,\tau)} , 
\end{equation}
where the vortex-vortex interaction $g_E(x,\tau)$ 
only depends on relative coordinates, and 
$\epsilon=\pi \hbar I/e$.  The contribution $F_G$ to  
the free energy due to regular configurations can be dropped, because
it does not acquire an imaginary part under the analytic continuation.
We  now perform the analytic continuation $\tau \rightarrow it$, 
resulting in [see Ref.~\cite{weissbook} for details]
\begin{equation}
{\rm Im} \, F = - \frac{Ly^2 c_s^2}{2\kappa^4} 
\int_{-L/2}^{L/2} dx \int_{-\infty}^{\infty} dt \,
e^{i \epsilon t -2 \mu g(x,t)} ,
\end{equation}
where $g(x,t)\equiv g_E(x,\tau\rightarrow it)$.
The rate $\Gamma(\epsilon)$ then follows 
for $L,L_T \gg \kappa$ but arbitrary $L/L_T$ in the form 
\[
\Gamma(\epsilon)=\frac{c_s^2 L y^2}{\kappa^4}
\int_{-L/2}^{L/2} dx \int_{-\infty}^\infty dt \, e^{i\epsilon t
- \mu [\tilde g(t+x/c_s)+ \tilde g(t-x/c_s)]} ,
\]
where 
\[
\tilde g(t) = \ln \left[(L_T/\kappa) \sinh(\pi T|t|)\right]
+ i (\pi/2) {\rm sgn}(t).
\]
Analyticity of $g_E(x,\tau)$ in the strip $0\leq\tau\leq 1/T$ 
also leads to the standard detailed balance relation 
\cite{weissbook},
\[
\Gamma(-\epsilon)=e^{- \epsilon/T} \Gamma(\epsilon) .
\]
In order to  explicitly evaluate the rate $\Gamma(\epsilon)$ for 
arbitrary $L/L_T$, we now replace the boundaries for the $x$-integral 
by a soft exponential cutoff, 
switch to integration variables
$t'=t-x/c_s$ and $t^{\prime \prime}=t+x/c_s$, 
and use the auxiliary relation
\[
e^{-c_s | t^{\prime\prime}- t'|/L}= \frac{c_s}{\pi L}
\int_{-\infty}^\infty ds \frac{e^{-is(t^{\prime\prime}-t')}}
{s^2+(c_s/L)^2}.
\]
The $t',t^{\prime\prime}$ time integrations then decouple,
and it is straightforward to carry them out.
Finally some algebra yields the linear resistance 
in the form
\begin{eqnarray}\label{resis}
\frac{R(T)}{R_Q} &=& \left(  \frac{\pi y \Gamma(\mu/2)}{ \Gamma(\mu/2+1/2)}
\right)^2 \frac{\pi L}{2\kappa} \left(\frac{L_T}{\kappa}\right)^{3-2\mu} 
\\ \nonumber
 &\times &  \int_{0}^\infty du \frac{2/\pi}{1+u^2}
\left|\frac{\Gamma(\mu/2+iu L_T/2L)} {\Gamma(\mu/2)} \right|^4 ,
\end{eqnarray}
in units of the resistance quantum $R_Q$ defined in Eq.~(\ref{contactres}).

For $L/L_T\gg 1$, the $u$-integral approaches unity, 
and hence $R\propto T^{2\mu-3}$,
while for $L/L_T\ll 1$, dimensional scaling arguments
give $R\propto T^{2\mu-2}$.  In Refs.~\cite{kociak,kasnew},
typical lengths were $L\approx 1~\mu$m, which puts one into the
crossover regime $L_T\approx L$. 
While the quoted power laws have already been reported for diffusive wires 
\cite{zaikin1}, Eq.~(\ref{resis}) describes the full
crossover for arbitrary $L/L_T$, and applies to
strongly correlated ladder compounds such as nanotube ropes.
It predicts that 
the transition gets significantly 
broader upon decreasing the number of tubes in the rope.
This is shown in Fig.~\ref{figevo}, where the theoretical
results for the resistance is
plotted for various $N$ at $\nu=1$.
Note that Eq.~(\ref{resis}) is a perturbative result
in the fugacity, and it is expected to break down close to $T_c$,
see below.
In the next section we directly compare Eq.~(\ref{resis}) to 
experimental data obtained by Kasumov {\sl et al.} \cite{kasnew}.

\section{Comparison to experimental data}\label{sec6}

\begin{figure}
\centerline{\epsfxsize=7cm \epsfysize=6cm \epsffile{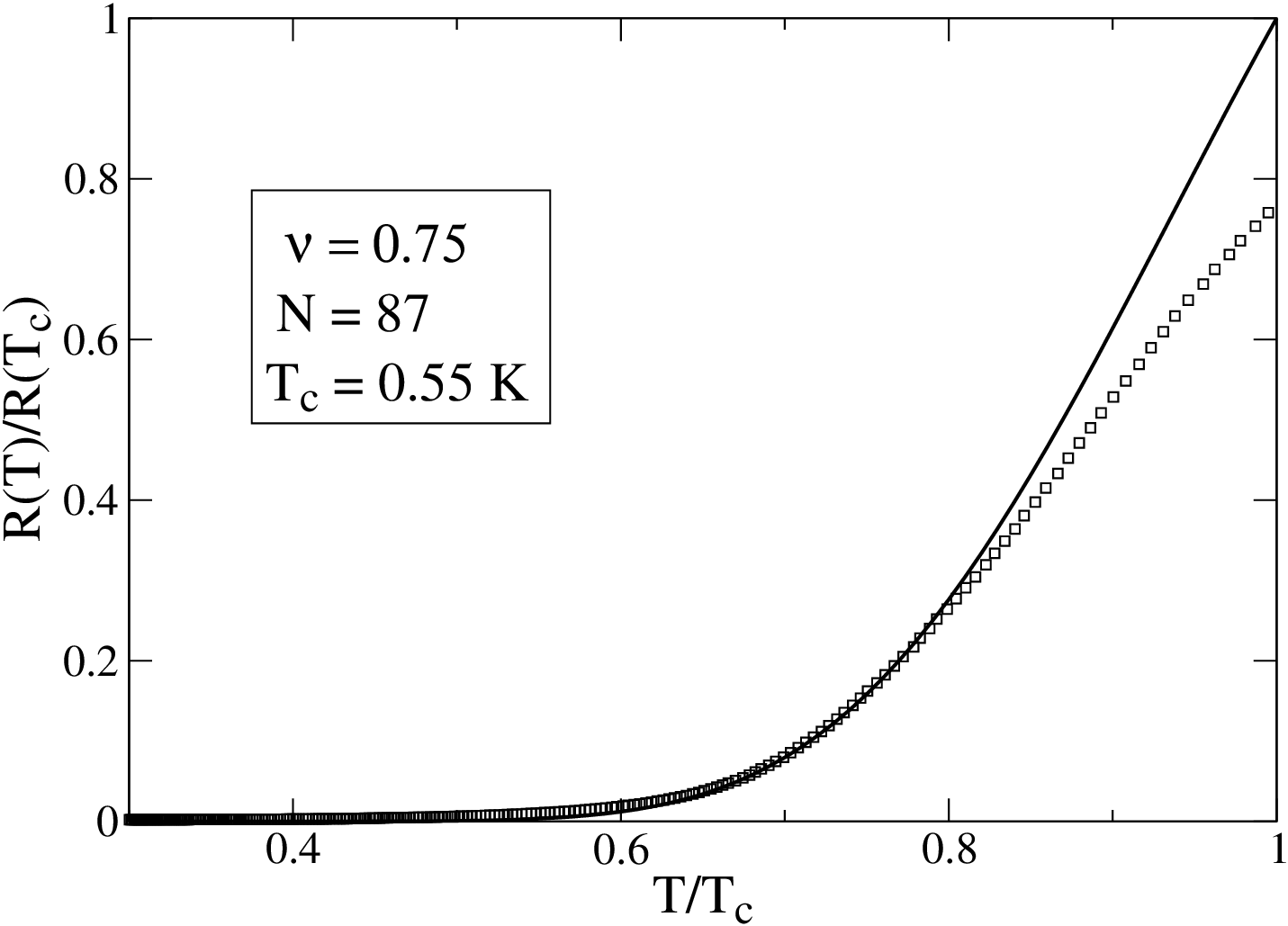}}
\caption{\label{figR2}  
Temperature dependence of the resistance below $T_c$ for 
superconducting rope 
$R2$ experimentally studied in Ref.~\cite{kasnew}. 
Open squares denote experimental data 
(with subtracted residual resistance), 
the curve is the theoretical result.
}
\end{figure}

Here we discuss how the prediction for the 
temperature-dependent resistance $R(T)$ below
$T_c$ as given in Eq.~(\ref{resis}) compares to the experimental
results for $R(T)$ published in Ref.~\cite{kasnew}. 
More aspects of this comparison will be given elsewhere \cite{next}.
The experimental data in Ref.~\cite{kasnew} were obtained from
two-terminal measurements of ropes suspended between 
normal electrodes. 
Due to the presence of the contacts, 
the residual resistance (\ref{contactres}) survives down to $T=0$ 
even when the rope exhibits a superconducting transition.
Extrapolation of experimental results for $R(T)$ yields $R_c$, which then 
allows to infer the number $N$ in the respective sample
 from Eq.~(\ref{contactres}). 
This resistance $R_c$ has to be subtracted from experimental
data to allow
for a comparison with Eq.~(\ref{resis}), where no contact resistance
 is taken into account.

\begin{figure}[t]
\centerline{\epsfxsize=7cm \epsfysize=6cm \epsffile{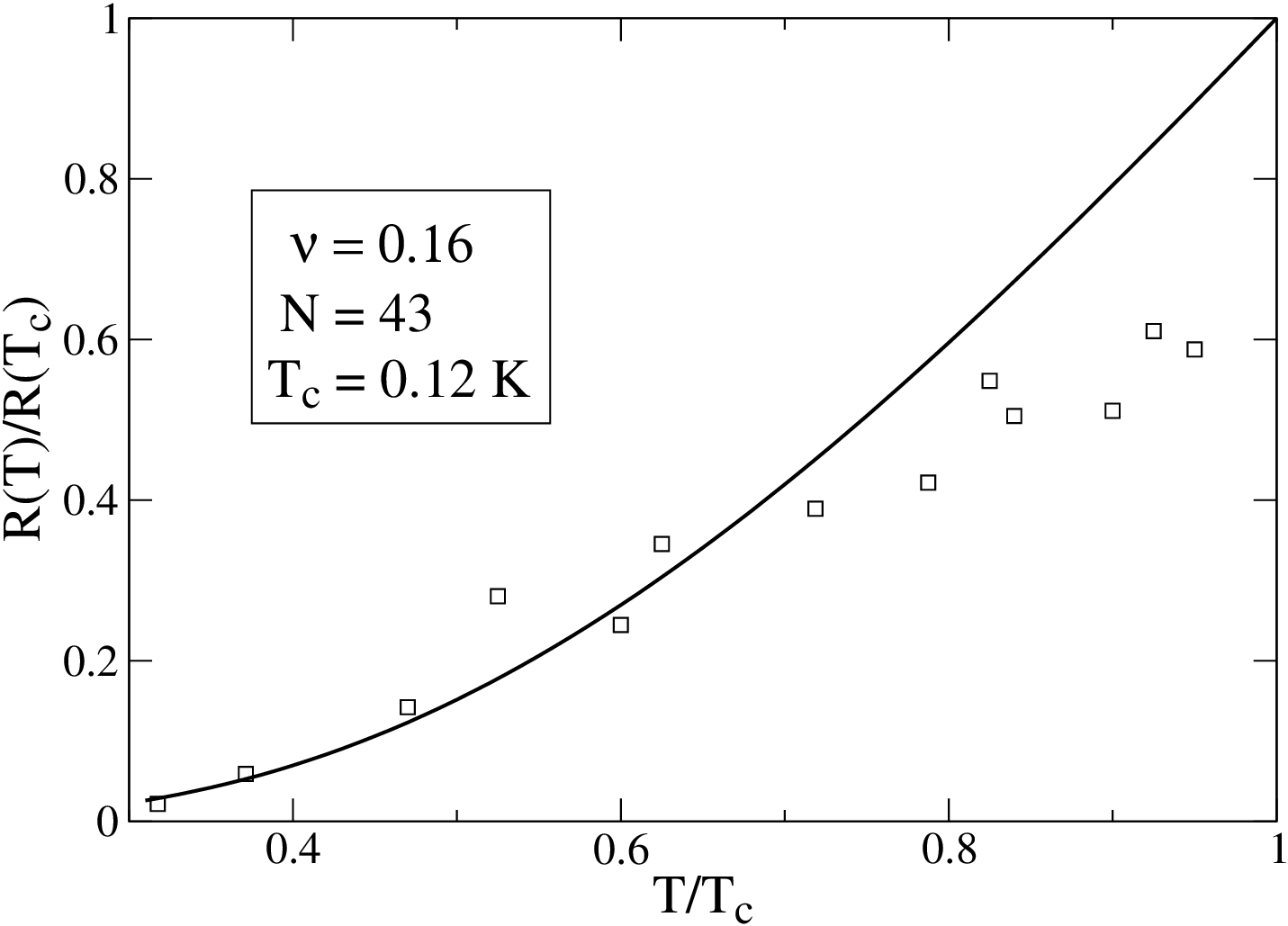}}
\caption{\label{figR4}  
Same as Fig.~\ref{figR2}, but for sample $R4$ experimentally
studied in Ref.~\cite{kasnew}.
}
\end{figure}

In Figs.~\ref{figR2} and \ref{figR4}, experimental
resistance curves (after this subtraction)
for the samples named $R2$ and $R4$ 
in Ref.~\cite{kasnew} are plotted versus the prediction 
of Eq.~(\ref{resis}). For sample 
$R2$, we find $R_c= 74~\Omega$ 
corresponding to $N_{R2}=87$, while for sample $R4$, the 
subtracted resistance is $R_c=150~\Omega$, 
leading to $N_{R4}=43$.  We then take these $N$ values 
when computing the respective
theoretical curves.
The experimentally determined
temperature $T^\ast$ locates the onset
of the transition \cite{kasnew}, and 
is identified with the true transition temperature $T_c$
in Eq.~(\ref{tc1}). It is therefore also not a free parameter. 
Note that thereby the eigenvalue $\Lambda_1$ of the Josephson
matrix has been determined.
In the absence of detailed knowledge about the
Josephson matrix, it is fortunate that our result 
for $R(T)/R(T_c)$ following from Eqs.~(\ref{resis}) 
and (\ref{mu1}) does not require more information about
$\Lambda$ besides the largest eigenvalue.
Given the estimate $g_{c+}=1.3$
\cite{ademarti}, the comparison of Eq.~(\ref{resis}) 
to experimental data then allows only one free fit 
parameter, namely $\nu$.  According to our discussion
in Sec.~\ref{sec41},  the fit is expected to yield  values $\nu \approx 1$.

The best fit to the low-temperature experimental curves for $R(T)$
 yields $\nu=0.75$ for sample $R2$, see Fig.~\ref{figR2},
and $\nu=0.16$ for sample $R4$, respectively.  The agreement between
experiment and theory is excellent for sample $R2$.  For sample $R4$,
the optimal $\nu$ is slightly smaller than expected, which indicates that
dissipative processes may be more important in that sample.
Nevertheless, for both samples, the low-temperature
resistance agrees quite well, with only one free fit parameter that
is found to be of order unity as expected.
Whereas the agreement between theoretical and experimental curves
appears then quite satisfactory in the low-temperature region, 
our predictions clearly deviate in the region near $T_c$. 
This is not surprising, because
our expression for $R(T)$ in Eq.~(\ref{resis}) 
is perturbative in the QPS fugacity. It is then expected to
break down close to $T_c$, where QPSs proliferate and the approximation
of a very dilute gas of QPS pairs, on which 
our calculation is based, is not valid anymore. 
As a consequence, also the saturation observed 
experimentally above  $T^\ast$ is not captured.

We note that it is an interesting challenge to compute the 
finite resistance in the normal phase at $T_c<T<T_c^0$,
where the saturation should be caused by
QPS and TAPS proliferation.  For temperatures $T>T_c^0$, superconducting
correlations can be neglected, and the resistance should then be dominated
by phonon backscattering and disorder effects.
Nevertheless, we believe that the 
agreement between the theoretical resistance result (\ref{resis}) 
and experimental data at low temperatures shown
in Figs.~\ref{figR2} and \ref{figR4}, 
given the complexity of this system,  is rather satisfactory.
More importantly, this comparison 
provides strong evidence for the presence of quantum phase
slips in superconducting nanotube ropes.  

\section{Conclusions} \label{sec7}

According to our discussion above,
the intrinsic superconductivity observed in ropes of 
SNWTs \cite{kasnew,kociak} represents a 
remarkable phenomenon, where it has been
possible to experimentally probe the extreme 1D
limit of a superconductor.
In this paper, we have formulated a theory for this phenomenon,
based on a model of metallic SWNTs with 
attractive intra-tube interactions
and arbitrary inter-tube Josephson couplings. 
The analysis of this model leads to an effective Ginzburg-Landau 
action, whose coefficients can be expressed in 
terms of parameters entering the 
microscopic description of the rope. In order
to get the correct low-energy dynamics, it is crucial 
to include  quantum fluctuations of the order parameter.
Based on the resulting low-energy action for the phase 
fluctuations, we have shown that quantum phase slips 
produce a depression of the critical temperature. 
More importantly, the temperature dependence of the linear resistance 
experimentally observed below the transition temperature 
can be accounted for by considering the underlying LL physics 
and the effect of quantum phase slips.
Despite some admittedly crude approximations, like the neglect of
intra-tube disorder and dissipation effects inside the vortex core, 
the comparison of 
experimental curves and theoretical predictions, 
in particular in the low-temperature region, strongly 
suggests that the resistive process is indeed dominated by 
quantum phase slips.
Our theory also suggests that, if repulsive Coulomb interactions 
can be efficiently screened off, superconductivity may survive down 
to only very few transverse channels in clean nanotube ropes.

\section*{Acknowledgments}

We thank H. Bouchiat and M. Ferrier
for providing the data shown in Figs.~\ref{figR2} and
\ref{figR4} and for many discussions. One of us (ADM) would like 
to thank them also for their warm hospitality 
during an extended stay in Orsay.
This work has been supported by the EU network DIENOW
and the SFB-TR 12 of the DFG.


\begin{references}

\bibitem{nts} 
C. Dekker, Physics Today {\bf 52(5)}, 22 (1999);
See also reviews on nanotubes in Physics World No.~{\bf 6} (2000).

\bibitem{nts2}
M. Dresselhaus, G. Dresselhaus and Ph. Avouris (Eds.),
Carbon Nanotubes, Topics in Appl. Physics {\bf 80}
(Springer Verlag 2001).

\bibitem{kociak}
M. Kociak, A.Yu. Kasumov, S. Gueron, B. Reulet, I.I. Khodos,  
Yu.B. Gorbatov, V.T. Volkov, L. Vaccarini, and H. Bouchiat,
Phys. Rev. Lett. {\bf 86}, 2416 (2001).

\bibitem{kasnew}
A. Kasumov, M. Kociak, M. Ferrier, R. Deblock, S. Gueron, B. Reulet, 
I. Khodos, O. Stephan, and H. Bouchiat, Phys. Rev. B {\bf 68}, 214521 (2003).

\bibitem{tang}
Z.K. Tang, L. Zhang, N. Wang, X.X. Zhang, G.H. Wen, G.D. Li, J.N. Wang, C.T. 
Chan, and P. Sheng, Science {\bf 292}, 2462 (2001).

\bibitem{kasumov}
A.Yu. Kasumov,  R. Deblock, M. Kociak, B. Reulet, H. Bouchiat, I.I. 
Khodos,  Yu.B. Gorbatov, V.T. Volkov, C. Journet, and M. Burghard,
Science {\bf 284}, 1508 (1999).

\bibitem{morpurgo}
A.F. Morpurgo, J. Kong, C.M. Marcus, and H. Dai,
Science {\bf 286}, 263 (1999).

\bibitem{lau}
C.N. Lau, N. Markovic, M. Bockrath, A. Bezryadin, and M. Tinkham,
Phys. Rev. Lett. {\bf 87}, 217003 (2001).

\bibitem{gonzalez1} J. Gonz{\'a}lez, 
Phys. Rev. Lett. {\bf 88}, 076403 (2002).

\bibitem{gonzalez2} J. Gonz{\'a}lez, 
Phys. Rev. B {\bf 67}, 014528 (2003).

\bibitem{ademarti} A.~De~Martino and R. Egger,
Phys. Rev. B {\bf 67}, 235418 (2003).

\bibitem{kane} A.A. Maarouf, C.L. Kane, and E.J. Mele,
Phys. Rev. B {\bf 61}, 11156 (2000).

\bibitem{schulz} H.J. Schulz and C. Bourbonnais,
  Phys. Rev. B {\bf 27}, 5856 (1983).

\bibitem{tinkham} M. Tinkham, {\sl Introduction to Superconductivity}, 2nd ed.
(McGraw-Hill, Inc., 1996).

\bibitem{zaikin1} 
  A.D. Zaikin, D.S. Golubev,  A. van Otterlo, and G.T. Zimanyi,
Phys. Rev. Lett. {\bf 78}, 1552 (1997).

\bibitem{zaikin2} D.S. Golubev and A.D. Zaikin,
Phys. Rev. B {\bf 64}, 014504 (2001).

\bibitem{blatter}
H.P. B{\"u}chler, V.B. Geshkenbein, and G. Blatter,
Phys. Rev. Lett.  {\bf 92}, 067007 (2004).

\bibitem{gonzalez3} J. Gonz{\'a}lez, Eur. Phys. J. B {\bf 36}, 317 (2003).

\bibitem{egger97} R. Egger and A.O. Gogolin,
Phys. Rev. Lett. {\bf 79}, 5082 (1997).

\bibitem{kane97}
  C. Kane, L. Balents, and M.P.A. Fisher,
Phys. Rev. Lett. {\bf 79}, 5086 (1997).

\bibitem{gogolin} A.O. Gogolin, A.A. Nersesyan, and A.M. Tsvelik,
{\sl Bosonization and Strongly Correlated Systems} (Cambridge University 
Press, 1998).

\bibitem{egger98} R. Egger and A.O. Gogolin,
Eur. Phys. J B {\bf 3}, 281 (1998).

\bibitem{carr} S.T. Carr and A.M. Tsvelik,
Phys. Rev. B {\bf 65}, 195121 (2002).

\bibitem{foot1}
Modes with $\Lambda_\alpha=0$ have to be excluded in the
transformation. All $\alpha$ 
summations have to be understood in this sense. 
Note that such modes never cause critical behavior in any case.

\bibitem{nagaosa1} N. Nagaosa,
{\sl Quantum Field Theory in Condensed Matter Physics}
(Springer Verlag, 1999).

\bibitem{mooij} 
  J. E. Mooij and G. Sch{\"o}n,
Phys. Rev. Lett. {\bf 55}, 114 (1985).

\bibitem{chaikin} P.M. Chaikin and T. Lubensky,
  {\sl Principles of Condensed Matter Physics} 
(Cambridge University Press, 2000).

\bibitem{langer}
J.S. Langer, Ann. Phys. (N.Y.) {\bf 41}, 108 (1967).

\bibitem{weissbook}
U. Weiss, {\sl Quantum Dissipative Systems}, 2nd. ed. 
(World Scientific, Singapore, 1999).

\bibitem{next}
M. Ferrier, A. De~Martino, A. Kasumov, R. Deblock, S. Gueron, R. Egger,
and H. Bouchiat, invited review, submitted to Solid State Communications.

\end{references}
\end{document}